\documentclass[twoside]{article}

\usepackage{amssymb}
\usepackage{icrctc07}

\title{Astrophysics Motivation behind the Pierre Auger Southern Observatory Enhancements}
\shorttitle{Astrophysics motivation behind Auger enhancements}
\authors{G. Medina Tanco$^{1}$, for the Pierre Auger Collaboration$^{2}$}
\shortauthors{Pierre Auger Collaboration}
\afiliations{$^{1}$Instituto de Ciencias Nucleares, Universidad
Nacional Autonoma de Mexico, A.P. 70-543, Ciudad Universitaria, D.F.
04510, Mexico. \\ $^{2}$Pierre Auger Observatory, Av. San
Mart\'{\i}n Norte 304, (5613) Malargue, Argentina}
\email{gmtanco@nucleares.unam.mx}

\abstract{ The Pierre Auger Collaboration intends to extend the
energy range of its southern observatory in Argentina for high
quality data from 0.1 to 3 EeV. The extensions, described in
accompanying papers, include three additional fluorescence
telescopes with a more elevated field of view (HEAT) and a nested
surface array with 750 and 433 m spacing respectively and additional
muon detection capabilities (AMIGA). The enhancement of the detector
will allow measurement of cosmic rays, using the same techniques,
from below the second knee up to the highest energies observed. The
evolution of the spectrum through the second knee and ankle, and
corresponding predicted changes in composition, are crucial to the
understanding of the end of Galactic confinement and the effects of
propagation on the lower energy portion of the extragalactic flux.
The latter is strongly related to the cosmological distribution of
sources and to the composition of the injected spectrum. We discuss
the science motivation behind these enhancements as well as the
impact of combined HEAT and AMIGA information on the assessment of
shower simulations and reconstruction techniques. }

\begin{document}

\maketitle


\section{Target energy interval}

\vskip -0.2cm

The Southern site of the Pierre Auger Observatory, which will be
completed in its original design during the next few months, has an
energy threshold of $\sim 3 \times 10^{18}$ eV and $\lesssim
10^{18}$ eV for surface (full efficiency) and fluorescence events
respectively. Two enhancements, AMIGA and HEAT \cite{AMIGA,HEAT},
are already planned in order to extend its energy threshold for high
quality data down to $10^{17}$ eV.

\begin{figure}[h]
\begin{center}
\includegraphics [width=0.48\textwidth]{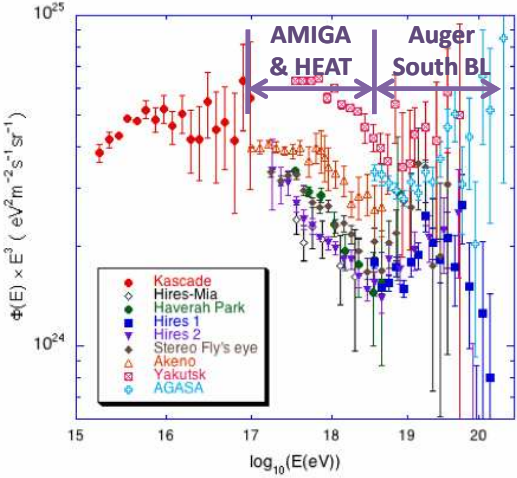}
\caption{ Cosmic ray energy spectrum and its main features: knee
(few PeV), second knee (~0.5 EeV) and ankle (EeV to few tens of
EeV). The energy regions covered by Auger baseline (BL) design and
that added by the enhancements AMIGA and HEAT are shown. Adapted
from ref. \cite{NIM_infill}.} \label{fig:espectrum_proposal}
\end{center}
\end{figure}

The baseline design of Auger is optimized for energies corresponding
to the middle of the ankle and upwards to the highest energies. The
enhancements, on the other hand, have as a prime objective to lower
the energy threshold of the detector down to $10^{17.0}$ eV (see,
figure \ref{fig:espectrum_proposal}).Such extension will allow the
complete inclusion of the ankle and the second knee inside the
observation range of Auger. This would also have the additional
advantage of adding an overlap with KASCADE Grande
\cite{KascadeGrande_design} which is of fundamental importance in
order to validate results.

The second knee has been observed in the vicinity of $4 \times
10^{17}$ eV by Akeno \cite{akeno_2ndknee}, Fly's Eye stereo
\cite{AbuZayyad2001}, Yakutsk \cite{Yakutsk_2ndKAnkleb} and HiRes
\cite{HiRes2004}. The physical interpretation of this spectral
feature is uncertain at present. It may be either the end of the
Galactic cosmic ray component, the pile-up from pair creation
processes due to proton interactions with the cosmic microwave
background radiation during propagation in the intergalactic medium
or a combination  of both effects.

The ankle, on the other hand, is a broader feature that has been
observed by Fly's Eye \cite{AbuZayyad2001} around $3 \times 10^{18}$
eV as well as by Haverah Park \cite{HP_Ankle} at approximately the
same energy. These results have been confirmed by Yakutsk
\cite{Yakutsk_2ndKAnkleb}, HiRes \cite{HiRes2004} and Auger in
hybrid mode \cite{auger_hybrid}. AGASA also observed the ankle, but
they locate it at a higher energy, around $10^{19}$ eV
\cite{Takeda2003}. Several physical interpretations are possible
which are intimately related with the nature of the second knee. The
ankle may be the transition point between the Galactic and
extragalactic components or the result of pair creation by protons
in the cosmic microwave background.

\section{Astrophysical characterization of the second knee and ankle}

\subsection{Observed composition} \label{sec:science_composition}

Certainly, the most relevant scientific result in the energy
interval previously defined will be the precise determination of the
chemical composition of the primary cosmic ray flux as a function of
energy.

Several techniques have been used to determine the composition of
cosmic rays in the energy region of the enhancements: depth of
maximum of the longitudinal distribution, $X_{max}$, fluctuations of
$X_{max}$, muon density, steepness of the lateral distribution
function, time profile of the signal and, in particular, rise time
of the signal, curvature radius of the shower front,
multi-parametric analysis, such as principal component analysis and
neural networks, etc.. Unfortunately, the different techniques give
conflicting results. The understanding of this inconsistencies is
critical to the understanding of the astrophysics of ultra high
energy cosmic rays.

At energies above few times $10^{16}$ eV, the flux is dominated by
heavy nuclei. These particles are of Galactic origin. What is being
detected is, very likely, the end of the efficiency of supernova
remnant shock waves as accelerators, since the Larmor radii or
characteristic diffusion scale lengths of the nuclei become
comparable to the radius of the remnants. If there are not more
powerful accelerators in the Galaxy \cite{Hillas_CRIS2004}, the
Galactic cosmic ray flux should continue to be dominated by iron
above $10^{17}$ eV and up to the highest energies produced inside
the Milky Way. At higher energies, the composition has been measured
by several experiments in the past, e.g., Haverah Park, Yakutsk,
Fly's Eye, HiRes-MIA prototype and HiRes in stereo mode and, more
recently, by Auger \cite{auger_xmax}.


The $X_{max}$ data suggest that, above $10^{16.6}$ eV, the
composition changes progressively from heavy to light. At the lower
limit of the target energy interval, the composition is heavy,
possibly iron dominated, in accordance to KASCADE results.
Nevertheless, at energies of $10^{19}$ eV, it is more consistent
with a flux dominated by lighter elements.

Despite the fact that there is a consensus among most of the
experiments about the reality of this smooth transition, there is no
consensus about the rate and extent to which the transition occurs.
In fact, the combined data from the HiRes-MIA prototype and HiRes in
stereo mode, point to a rapid transition from heavy to light
composition between $10^{17}$ and  $10^{18}$ eV \cite{HiResMIAXmax}.
Beyond that point, the composition would remain light and constant.


The later scenario, however, is not supported by the data of other
experiments. Haverah Park, for example, shows a predominantly heavy
composition up to $10^{18}$ eV, followed by an abrupt transition to
lighter values compatible with HiRes stereo at around $10^{18}$ eV.
Volcano Ranch, even though there is a single experimental point, is
compatible with a heavy composition still at $10^{18}$ eV, somewhat
in accordance to Haverah Park data. Akeno (A1), on the other hand,
is consistent with a continuation of the gradual transition from the
second knee all across the ankle up to at least $10^{19}$ eV. The
emerging picture is one of great uncertainty, which has deep
practical implications and imposes severe limitations on theoretical
efforts.


\subsection{Cosmic ray propagation}

At the energies of the enhancements, $10^{17}-10^{19}$ eV, there is
a change in the origin, acceleration and propagation regime of
primary cosmic rays. At lower energies, the Galaxy is undoubtedly
the source of cosmic rays. Several acceleration mechanisms are
certainly at play but it is widely expected that the dominant one is
first order Fermi acceleration at the vicinity of supernova remnant
shock waves. Nevertheless, theoretically, these Galactic
accelerators should become inefficient between $\sim 10^{17}$ and
$\sim 10^{18}$ eV. This upper limit could be extended to $\sim
10^{19}$ eV if additional mechanisms were operating in the Galaxy,
e.g., compact supernova remnants, spinning inductors associated with
compact objects, etc..

At these energies, particles also start to be able to travel from
the nearest extragalactic sources in less than a Hubble time.
Consequently, at some point above $10^{17.5}$ eV a sizable cosmic
ray extragalactic component should be detectable, probably becoming
dominant above $10^{19}$ eV. Therefore, it is expected that the
cosmic ray flux detected between the second knee and the ankle of
the spectrum is a mixture of the Galactic and extragalactic
components, highlighting the astrophysical richness and complexity
of the region.

The Galaxy is a magnetized medium, with a field structured on scales
of kpc and typical intensities of the order of a few micro Gauss.
This transforms the Galaxy in an efficient confinement region for
low energy charged particles. The confinement region is a flattened
disk of approximately 20 kpc of radius and thickness of the order of
a few kpc. The Larmor radius of a nucleus of charge Ze can be
conveniently parameterized as $r_{L,kpc} \approx \frac{1}{Z} \times
\left( E_{EeV} / B_{\mu G} \right)$ where $E_{EeV}$ is the energy of
the particle in units of $10^{18}$ eV and $r_{L,kpc}$ is expressed
in kpc. Protons with energies $\gtrsim 10^{17}$ eV have gyroradii
comparable or larger than the transverse dimensions of the effective
confinement region and, therefore, can easily escape from the
Galaxy. On the other end of the mass spectrum, just the opposite
occurs for iron nuclei that, even at energies of the order of
$10^{19}$ eV, have gyroradii $< 10^{2}$ pc and must be effectively
confined inside the magnetized interstellar medium.

Therefore, along the energy region targeted by the enhancements,
extending from the second knee up to almost the end of the ankle,
all nuclei from p to Fe, i.e. $1 < Z < 26$, experience a transition
in their propagation regime inside the interstellar medium changing
gradually from diffusive to ballistic as the energy increases.

Large statistics and a thorough knowledge of the acceptance across
the ankle and second knee, should make possible to measure large
scale anisotropies potentially associated with the
galactic/extragalactic transition.

Furthermore, extragalactic cosmic rays start to penetrate inside the
Galactic confinement region. However, extragalactic particles must
first be able to reach us from the nearest Galaxies in less than a
Hubble time.

A crude approximation to these effect can be made by assuming a Bohm
diffusion coefficient, which implies a travel time for extragalactic
cosmic rays $\tau_{Myr} \thickapprox 10 \times D_{Mpc}^{2} \times Z
\times \left(  \frac{B_{nG}}{E_{EeV}}  \right) \;$. This shows that
there is a rather restrictive magnetic horizon. Basically, no
nucleus with energy smaller than $10^{17}$ eV is able to arrive from
regions external to the local group ($D \sim 3$ Mpc) if indeed the
intergalactic fields have $nG$ strengths. Taking as a minimum
characteristic distance $D = 10$ Mpc, only protons with $E > 2
\times 10^{17}$ eV, or Fe nuclei with $E > 5 \times 10^{18}$ eV are
able to reach the Galaxy in less than a Hubble time.

Therefore, it is at the energies of the second knee and the ankle
that different nuclei start to arrive from the local universe.
Concomitantly, at these same energies, the magnetic shielding of the
Galaxy becomes permeable to these nuclei, allowing them to get into
the interstellar medium and, eventually, to reach the solar system.
Effectively, the energy interval from $\sim 2 \times 10^{17}$ to
$10^{19}$ eV is the region of mixing between the Galactic and
extragalactic components of cosmic rays.

At energies smaller than $\sim 10^{19.2}$ eV the dominant process is
the photo-production of electron-positron pairs in interactions with
the CMBR. In fact, the structure of the ankle can be explained
exclusively as a result of pair photo-production by nucleons
travelling cosmological distances between the source and the
observer \cite{ankle_by_pairprod_Berezinsky}.

The enhancements are designed to operate in the energy region where
the superposition of the Galactic and extragalactic spectra takes
place. This is a theoretically challenging region where the smooth
matching of the two rapidly varying spectra has yet to be explained.
It must be noted that, even if the shape of the spectrum is
important, it is by far insufficient to decipher the underlying
astrophysical model. The variation of the composition as a function
of energy turns then into the key to discriminate both fluxes and to
select among a variety of theoretical options.

There is an additional problem at very high energies above the
threshold for pion photo-production. When the statistics are low,
different astrophysical scenarios can produce energy spectra
experimentally indistinguishable at very high energy
\cite{ModelDegeneracy_Stanev}. This degeneracy, beyond few times
$10^{19}$ eV, can only be broken with supplementary information
coming either from higher energy neutrinos or from composition
measurements at lower energies, in the region of the second knee and
ankle.

\section{Composition at injection and the ankle}

Power law spectra injected at cosmological sources with different
compositions can produce experimentally very similar spectra at the
highest energies. Nevertheless, they can be distinguished at smaller
energies in the ankle region.

In particular, a purely protonic flux can reproduce the ankle
feature solely as an effect of photo production of electron-positron
pairs in interactions with the CMBR. In this case, the transition
between the Galactic and extragalactic fluxes must be located at the
second knee or very near to it.

In the case of a heavier mixed composition injected at the sources,
the ankle must be the result of the competition between the Galactic
and extragalactic spectra. Moreover, the composition will be a
strong function of energy inside this interval, giving an additional
tool to assess details of the astrophysical model.

\section{Conclusions}

Auger, in its baseline design, is an excellent instrument for the
determination of cosmic ray observables at the highest energies,
with particular emphasis in the resolution of the GZK controversy,
the search for extragalactic point sources, and the discrimination
between bottom-up and top-down production models. However, it has
become increasingly clear that further discrimination between
astrophysical models requires the knowledge of the evolution of the
cosmic ray composition along the transition region starting at the
second knee and encompassing the ankle, i.e. from few times
10$^{17}$ eV to $\gtrsim 10^{19}$ eV. This energy range is not
thoroughly covered by the present configuration of the Auger
Observatory, which is fully efficient only above $3\times 10^{18}$
eV.

The determination of the composition and its energy dependence
inside the transition region is a primordial objective of the
enhancements. Working in conjunction with the baseline Auger SD and
FD detectors, the enhancements will aid in a fundamental way to our
understanding of ultra-high energy cosmic rays in its astrophysical
context.


\bibliographystyle{plain}

\end{document}